\documentclass[twocolumn]{aastex631}

\usepackage[T1]{fontenc}
\DeclareRobustCommand{\VAN}[3]{#2}
\let\VANthebibliography\thebibliography
\def\thebibliography{\DeclareRobustCommand{\VAN}[3]{##3}\VANthebibliography}

\usepackage{graphicx}	
\usepackage{amsmath}	
\usepackage{amssymb}	
\usepackage{acronym} 

\usepackage{xspace}
\usepackage{multirow}
\newcommand{\monei}{\ensuremath{m_{1,\rm{i}}}\xspace}
\newcommand{\mtwoi}{\ensuremath{m_{2,\rm{i}}}\xspace}

\newcommand{\ai}{\ensuremath{a_{\rm{i}}}\xspace}
\newcommand{\qi}{\ensuremath{q_{\rm{i}}}\xspace}
\newcommand{\Zi}{\ensuremath{Z_{\rm{i}}}\xspace}
\newcommand{\vk}{\ensuremath{v_{\rm{k}}}\xspace}
\newcommand{\thetak}{\ensuremath{{\theta}_{\rm{k}}}\xspace}

\newcommand{\ei}{\ensuremath{{e}_{\rm{i}}}\xspace}

\newcommand{\Msun}{\ensuremath{\,\rm{M}_{\odot}}\xspace}
\newcommand{\Zsun}{\ensuremath{\,\rm{Z}_{\odot}}\xspace}
\newcommand{\kms}{\ensuremath{\,\rm{km}\,\rm{s}^{-1}}\xspace}
\newcommand{\AU}{\ensuremath{\,\mathrm{AU}}\xspace}

\newcommand{\alphaCE}{\ensuremath{\alpha_{\rm{CE}}}\xspace}
\newcommand{\sigmaRMS}{\ensuremath{\sigma_{\rm{rms}}^{\rm{1D}}}}

\acrodef{DCO}{double compact object}
\acrodef{BNS}{binary neutron star}
\acrodef{BBH}{binary black hole}
\acrodef{BHNS}{black hole--neutron star}
\acrodef{NS}{neutron star}
\acrodef{BH}{black hole}
\acrodef{GW}{gravitational wave}
\acrodef{SFR}{star formation rate}
\acrodef{GW}{gravitational wave}
\acrodef{SFR}{star formation rate}
\acrodef{SFRD}{star formation rate density}
\acrodef{SN}{supernova}
\acrodef{SNe}{supernovae}
\acrodef{CCSNe}{core collapse supernovae}
\acrodef{RMP}{remnant mass prescription}
\acrodef{CE}{common envelope}
\defcitealias{Boesky_2023_GW_paper}{Paper II}

\begin{document}
\title{Investigating the Cosmological Rate of Compact Object Mergers from Isolated Massive Binary Stars}
\author[0009-0005-9830-9966]{Adam P. Boesky}
\affiliation{Center for Astrophysics \textbar{} Harvard \& Smithsonian,
60 Garden St., Cambridge, MA 02138, USA}
\author[0000-0002-4421-4962]{Floor S. Broekgaarden}
\affiliation{Center for Astrophysics \textbar{} Harvard \& Smithsonian,
60 Garden St., Cambridge, MA 02138, USA}
\affiliation{AstroAI at the Center for Astrophysics \textbar{} Harvard \& Smithsonian,
60 Garden St., Cambridge, MA 02138, USA}
\affiliation{Simons Society of Fellows, Simons Foundation, New York, NY 10010, USA}
\affiliation{Department of Astronomy and Columbia Astrophysics Laboratory, Columbia University, 550 W 120th St, New York, NY 10027, USA}
\affiliation{William H. Miller III Department of Physics and Astronomy, Johns Hopkins University, Baltimore, Maryland 21218, USA}
\author[0000-0002-9392-9681]{Edo Berger}
\affiliation{Center for Astrophysics \textbar{} Harvard \& Smithsonian,
60 Garden St., Cambridge, MA 02138, USA}
\affiliation{The NSF AI Institute for Artificial Intelligence and Fundamental Interactions}
\begin{abstract}

Gravitational wave detectors are observing compact object mergers from increasingly far distances, revealing the redshift evolution of the binary black hole (BBH)---and soon the black hole-neutron star (BHNS) and binary neutron star (BNS)---merger rate.
To help interpret these observations, we investigate the expected redshift evolution of the compact object merger rate from the isolated binary evolution channel.
We present a publicly available catalog of compact object mergers and their accompanying cosmological merger rates from population synthesis simulations conducted with the COMPAS software.
To explore the impact of uncertainties in stellar and binary evolution, our simulations use two-parameter grids of binary evolution models that vary the common-envelope efficiency with mass transfer accretion efficiency, and supernova remnant mass prescription with supernova natal kick velocity, respectively. 
We quantify the redshift evolution of our simulated merger rates using the local ($z\sim 0$) rate, the redshift at which the merger rate peaks, and the normalized differential rates (as a proxy for slope).
We find that although the local rates span a range of $\sim 10^3$ across our model variations, their redshift-evolutions are remarkably similar for BBHs, BHNSs, and BNSs, with differentials typically within a factor $3$ and peaks of $z\approx 1.2-2.4$ across models.
Furthermore, several trends in our simulated rates are correlated with the model parameters we explore.
We conclude that future observations of the redshift evolution of the compact object merger rate can help constrain binary models for stellar evolution and gravitational-wave formation channels.
\end{abstract}
%

\section{Introduction} \label{sec:intro}

Observations of \acp{GW} from compact object mergers are revolutionizing our understanding of stellar mass \acp{BH} and \acp{NS} across cosmic time. 
To date, data taken with the detector network consisting of Advanced LIGO \citep{TheLIGOScientificDetector:2014jea} Advanced Virgo \citep{Virgo:2015}, and KAGRA \citep{Akutsu:2021} include on the order of $100$ statistically significant \ac{GW} observations of \acp{BBH} out to redshifts  $z\sim1.5$ \citep[e.g.,][]{Abbott:2021GWTC3, Nitz:2023-4-OGC, Venumadhav:2019, Venumadhav:2020, 2019PhRvD.100b3007Z, Olsen:2022, mehta2023new, Wadekar:2023}. 
These observations already probe the BBH merger rate as a function of redshift \citep[e.g.,][]{GWTC-3_population_inference,  Nitz:2023-4-OGC, Callister:2023, Payne:2023, Ray:2023}, and future observing runs equipped with technological upgrades such as O4, O5, and $A^{\#}$ are expected to increase the detection volume for stellar mass \acp{BBH} out to redshift $z\sim 2$ \citep[e.g.,][]{Baibhav:2019, Voyager:2020, Gupta:2023}.  
Moreover, next-generation detectors like the Einstein Telescope and Cosmic Explorer are expected to make $\gtrsim 100,000$ detections annually from \ac{BNS} and \ac{BHNS} mergers out to redshifts $z \gtrsim 2$ and \ac{BBH} mergers out to redshift $\gtrsim 10$.
The observational capacities of these detectors will allow us to measure the redshift distribution of mergers to within percent level precision  \citep{EinsteinTelescope:2010Punturo, EinsteinTelescope:2012Sathyaprakash, EinsteinTelescope:2020Maggiore, CosmicExplorer:2019reitze, CosmicExplorer:2021evans, Borhanian:2022arXiv220211048B,  Iacovelli:2022, Singh_2022, Gupta:2023}. 

\begin{figure*}
  \includegraphics[width=\textwidth]{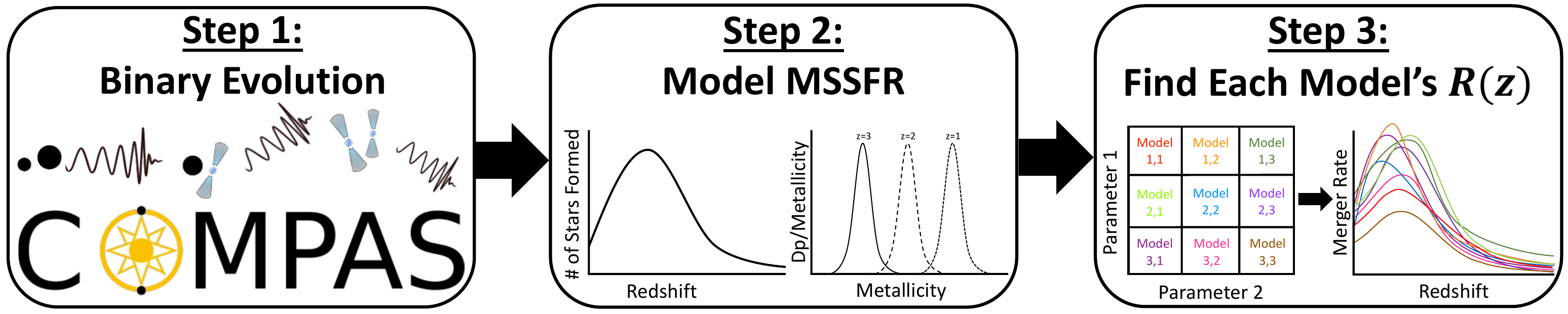}
  \caption{Schematic overview of the method used in this paper. Step 1: Simulate the evolution of binaries with each combination of parameters from the two-parameter grids (Table~\ref{tab:COMPAS_grid}) using the COMPAS binary population synthesis code. 
  Step 2: Distribute binaries from simulations across redshift using a metallicity-specific star formation rate. 
  Step 3: Infer the \ac{BBH}, \ac{BHNS}, and \ac{BNS} merger rates from the simulation results of each parameter combination in the two-parameter grids.}
  \label{fig:methods}
\end{figure*}

To realize the full potential of these \ac{GW} observations, we use theoretical models of binary evolution to infer their formation channels and learn about the underlying physical processes that lead to compact object mergers  \citep[e.g.,][]{Zevin:2021, Mapelli:2021review, mandelfarmer:2022}. 
Thus far, the majority of literature has attempted to compare simulated merger rates to the observed \textit{local} ($z\sim0$) merger rates, but the local rates alone do not provide enough information to distinguish formation pathway contributions due to the many poorly-constrained parameters in population synthesis simulations \citep[][]{MandelBroekgaarden:2022}.
There has therefore been increasing interest in investigating the properties and rates of \ac{DCO} mergers \textit{as a function of redshift} because of the information that they encode about formation channels \citep[e.g.,][]{Ng_2021, Zevin:2021, Singh_2022, vanSon:2022}.

To explore the cosmological merger rates, this study investigates the redshift distribution of \ac{DCO} mergers created by the isolated binary evolution channel, in which \ac{GW} sources form from pairs of massive stars. 
Our analysis is more in depth than most previous studies in three ways: (i) we analyze the \ac{BBH}, \ac{BHNS}, and \ac{BNS} merger rates simultaneously, (ii) we explore the impact of uncertain populations synthesis parameters in tandem using two grids of simulations with varying assumptions for the mass transfer, common envelope, and supernova physics, and (iii) we calculate summary statistics, such as the `differential merger rates' across several redshift bins to efficiently analyze the impact of different massive binary evolution uncertainties on the expected distribution of mergers.
Our simulations are publicly available at \url{https://gwlandscape.org.au/compas/}.

\section{Methods} \label{sec:methods}

We calculate the merger rates of simulated \ac{BBH}, \ac{BHNS}, and \ac{BNS} systems formed by isolated massive binary stars in a three-step process shown in Figure~\ref{fig:methods} and summarized below.

\begin{table*}
    \caption{Overview of the models explored in this study. For the remnant mass prescriptions we use the `delayed' and `rapid' prescriptions from \citet{Fryer_2012} and the `stochastic' prescription from \citet{Mandel_2020}.}
    \label{tab:COMPAS_grid}
    \resizebox{\linewidth}{!}{%
    \hspace{-1.6cm}
    \begin{tabular}{lllll}
    \hline  \hline
    Grid label (dimensions)		& Parameters & Values    & Changed physics    \\ \hline  \hline
    \multirow{2}{*}{\ \raisebox{-0.4 cm}{\huge A} (4x3)} & $\alphaCE$ & $ [0.1, 0.5, 2.0, 10.0]$  & Common envelope ejection efficiency\\
     & $\beta$ & $[0.25, 0.5, 0.75]$ & Mass transfer accretion efficiency & 
    \\
    \hline \hline
    \multirow{2}{*}{\ \raisebox{-0.4 cm}{\huge B} (3x3)} & Supernova & [delayed, rapid, Mandel \& M\"uller]   & SN remnant mass  prescription\\
     & $\sigmaRMS$ & $[30, 265, 750]\, \rm{km}\ \rm{s}^{-1}$ & 1d rms natal kick velocity & 
    \\
    \hline \hline
    \end{tabular}%
    }
\end{table*}

\subsection{Population Synthesis Simulations}

We use the COMPAS\footnote{Compact Object Mergers: Population Astrophysics and Statistics, \url{https://compas.science}.} suite to rapidly evolve large populations of stellar binaries, a fraction of which create compact objects and merge \citep{COMPAS_2022}.
COMPAS is built on the single star evolution analytic fitting formulae by \citet{Hurley_2000, Hurley_2002} which are based on single star evolution tables from \cite{Pols_1998} and earlier work from \citet{Eggleton_1989} and \citet{Tout_1996}; it parameterizes and approximates stellar evolution and binary interaction in order to rapidly ($\lesssim 1$ sec) compute the evolution of binary systems.
We ran population synthesis simulations over two dimensional grids of model parameters to explore the correlated impact of binary physics uncertainties.
We created two grids, which we will refer to as grid A and grid B, and which we summarize below and in Table~\ref{tab:COMPAS_grid}.

In grid A, we vary \ac{CE} efficiency and mass transfer efficiency. These two parameters are of interest because they have been the focus of several prior studies, are highly uncertain, and have been demonstrated to significantly impact populations synthesis outcomes \citep{VignaGomez_2018, Bavera_2021, Dorozsmai:2024, Santoliquido_2021, Broekgaarden_2022}. 
\ac{CE} phases are defined by dynamically unstable mass transfer in which on of the companions' envelope engulfs the other, causing drag and tightening the binary.
COMPAS parameterizes the \ac{CE} phase with the `$\alphaCE - \lambda$' formalism (introduced by \citet{Webbink_1984} and \citet{deKool_1990}), where $\alphaCE$ determines the fraction of orbital energy that binaries expend to eject their CEs.
For grid A we choose a range of values representative of the literature: $\alphaCE = 0.1,\ 0.5,\ 2.0,\textrm{ and }10.0$ \citep{Santoliquido_2022, Broekgaarden_2021, Neijssel_2019, van_Son_2022}, and we fix $\lambda$ to $\lambda_{\textrm{Nanjing}}$ from the fit in \citet{Xu_2010a, Xu_2010b}. 
The mass transfer efficiency parameter is $\beta = \Delta M_{\rm acc}/M_{\rm donor}$ where $\Delta M_{\rm donor}$ and $\Delta M_{\rm acc}$ are the changes in the mass of the donor and accretor stars, respectively, during stable transfer.  
We use three values, $\beta = 0.25,\ 0.5$, and $0.75$.

In grid B, we vary the \ac{CCSNe} natal kick velocity and the \ac{SN} \ac{RMP}.
Our simulations give \ac{SNe} a kick with velocity $v_k$ drawn from a Maxwell-Boltzman distribution with dispersion $\sigmaRMS$.
We explore $\sigmaRMS = 30,\ 265,\textrm{ and } 750 \ \textrm{km s}^{-1}$ for \ac{CCSNe}. Higher dispersion leads to faster kicks, which studies have found to be proportional to the amount of ejecta from \ac{SNe}.
We therefore choose these dispersion values to approximate having weak kicks ($\sigmaRMS = 30$), commonly-used moderate kicks ($\sigmaRMS = 265$ from \citet{Hobbs_2005}), and strong kicks ($\sigmaRMS = 750$).
\ac{RMP} maps objects' carbon-oxygen core mass to a remnant mass after \ac{SN}, and is largely responsible for determining if stars become \acp{NS} or \acp{BH}.
In grid B, we adopt three \acp{RMP} for \ac{CCSNe}: ``delayed'' \citep{Fryer_2012}, ``rapid'' \citep{Fryer_2012}, and ``stochastic'' \citep{Mandel_2020}.
The rapid prescription assumes that \ac{SN} explosions occur within $250$ ms as opposed to the longer duration in the delayed model; the rapid model reproduces a mass gap between \acp{NS} and \acp{BH} \footnote{Theoretical and observational studies predicted a gap between the masses of \acp{BH} and \acp{NS} in the $3$--$5\ M_\odot$ range, however \citet{LIGO_mass_gap} recently reported a merger with a component mass of $2.5$--$4.5\ M_\odot$. The mass gap remains a topic of debate.}.
The ``stochastic'' model, on the other hand, enables \ac{NS} and \ac{BH} formation in multiple regions of the parameter space.
Kick velocity is associated with mass ejecta, which in turn reduces the remnant mass as remnants accumulate ejecta through "fallback" driven by their gravitational pull.

For each pairing of parameter values in our grids, we evolve $20$ million binaries with initial stellar masses drawn from the \citet{Kroupa_2001} initial mass function in the $5-150 \Msun$ range. For all parameters not varied by the grids in this study, we use the default values from COMPAS \citep{COMPAS_2022} which are listed in appendix Table \ref{tab:COMPAS_fiducial}.

\subsection{Calculating the Merger Rate}
We calculate the cosmological merger rates of compact objects following the methodology in \citet{COMPAS_2022}. The merger rate measured by a comoving observer at a merger time $t_m$ since the Big Bang for a binary consisting of components with masses $M_1,$ and $M_2$ is
\begin{equation}
\begin{split}
    &\mathcal{R}_{\rm{merge}}(t_m, M_1, M_1) \equiv  \frac{\textrm{d}^4N_{\textrm{merge}}}{\textrm{d}t_m\textrm{d}V_c\textrm{d}M_1\textrm{d}M_2} (t_\textrm{m}, M_1, M_2)                       
    \\ &= \int \textrm{d}Z_i \int_0^{t_m}\textrm{d}t_{\textrm{delay}}\mathcal{S}(Z_i, z(t_\textrm{form} = t_m  - t_\textrm{delay})) \times \\ &\frac{\textrm{d}^4N_{\textrm{form}}}{\textrm{d}M_\textrm{SFR}\textrm{d}t_\textrm{delay}\textrm{d}M_1\textrm{d}M_2}(Z_i, t_\textrm{delay}, M_1, M_2),
\end{split}
\label{eq:merger_rate}
\end{equation}
where $N_{\rm{merger}}$ is the number of systems merged, $N_{\rm{form}}$ is the number of systems formed, $V_c$ is the comoving volume, $t_\textrm{delay}$ is the time between the formation and merger of the binary, $Z_i$ is the birth metallicity of the components, $M_{\rm{SFR}}$ is a unit of star forming mass, and we convolve the metallicity-specific star formation rate density $\mathcal{S}(Z_i, z(t_\textrm{form}))$ with the formation yield.

\subsection{Metallicity-Specific Star Formation History}

In order to model $S(Z_i, z(t_{\rm{form}}))$, which describes star formation history as a function of initial redshift and metallicity, we follow \citet{COMPAS_2022, Broekgaarden_2019}:
we multiply the \ac{SFRD} by a metallicity probability density function
\begin{equation}
\begin{split}
   \mathcal{S}(Z_i, z_{form})  &= \frac{\textrm{d}^3M_{\textrm{SFR}}}{\textrm{d}t_s\textrm{d}V_c\textrm{d}Z_i}(z_{form}) \\ 
    &= \frac{\textrm{d}^2M_{\rm{SFR}}}{\textrm{d}t_s\textrm{d}V_c}(z_{form})\times\frac{\textrm{d}P}{\textrm{d}Z_i}(z_{form})
\end{split}
\end{equation}
where $z_{\rm{form}}$ is the redshift at which \acp{DCO} form and $t_s$ is the time in the merger's source frame. We obtain the metallicity density function $\frac{\textrm{d}P}{\textrm{d}Z_i}(z_{\rm{form}})$ by convolving the number density of galaxies per logarithmic mass bin (GSMF) and the mass-metallicity relation.

%

For this study we use the SFRD fit from \citet{Madau_Fragos_2017}, which is an update from the earlier work \citet{Madau_Dickinson_2014}. 
We adopt the GSMF from \cite{Panter_2004} which is a standard Schechter fit based on the Sloan Digital Sky Survey data. We use the MZR from  \cite{Ma_2016}, which was derived using high-resolution cosmological zoom-in simulations from \citep{Hopkins_2014}. 
Models for $S(Z, z)$ are another source of high uncertainty which impacts merger rate approximation \citep{chruslinska_2022, Broekgaarden_2022, vanSon:2022}.
Evaluating the impact of the \ac{SFRD} on the merger rate will be a substantial effort that we leave for future studies.

\subsection{Quantifying the Merger Rate Redshift Evolution}
To understand how parameter variations impact the redshift evolution of the merger rate, it is helpful to use summary statistics for the $z$-distribution and behavior of $\mathcal{R}_{\rm{merge}}(z)$.
We quantify the merger rate $z$-evolution by calculating the relative differential rates
\begin{equation}
    \frac{\Delta \mathcal{R}_{\rm{merge}}}{\Delta z} (z_{\rm{min}}, z_{\rm{max}}) = \frac{\mathcal{R}_{\rm{m}}(z_{\rm{max}}) - \mathcal{R}_{\rm{m}}(z_{\rm{min}})}{(z_{\rm{max}} - z_{\rm{min}})\int_0^{\infty} \mathcal{R}_{\rm{m}}(z)dz}
    \label{eq:differential}
\end{equation}
where $\mathcal{R}_{\rm{m}}(z)$ is shorthand for $\mathcal{R}_{\rm{merge}}(z)$, $z_{\rm{min}}$ and $z_{\rm{max}}$ set the bounds for the differential rate, and the integral in the denominator is used to scale the rates such that the differential values describe redshift evolution instead of the number of mergers.
We choose this metric because it represents the slopes (relative increase) of the merger rate for a given redshift bin.

In this study, we calculate the differentials for the redshift ranges $[0, 1]$, $[1, z_{\rm{peak}}]$, $[z_{\rm{peak}}, z_{\rm{peak}} + 1]$, $[z_{\rm{peak}} + 1 , 9]$, where $z_{\rm{peak}}$ is the redshift of the peak. 
We select these ranges because (i) our rates monotonically increase and then decrease before and after the peak, respectively, (ii) it allows for breaks in the slope before and after the peak (as has been suggested by observational studies such as \citet{Callister:2023, Payne:2023}), and (iii) the slope of $\mathcal{R}_{\rm{merge}}(z)$ far from the peak is highly-linear. 
When we discuss the redshift evolution of the merger rate in Section~\ref{sec:results}, we discuss the differentials as well as the intrinsic merger rate, $\mathcal{R}_0 \equiv \mathcal{R}_{\rm{merge}}(0)$, and the redshift of the merger rate peak, $z_{\rm{peak}}$.


\section{Results} \label{sec:results}

\begin{figure*}
  \centering
  \includegraphics[width=\linewidth]{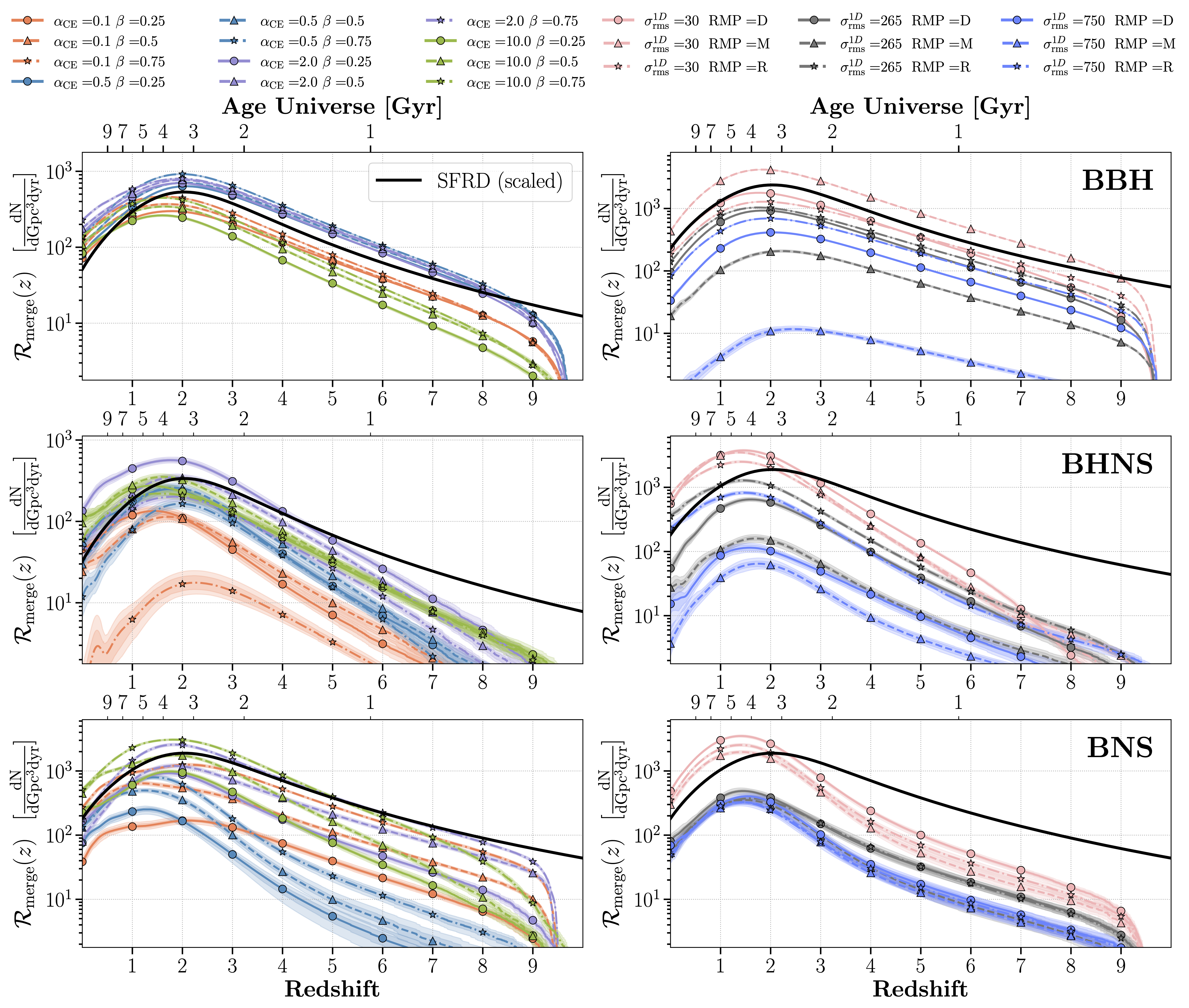}
  \caption{The BBH (top), BHNS (middle), and BNS (bottom) merger rates as a function of redshift for our simulations. 
  The left column shows models from grid A, where we vary the common-envelope efficiency $\alphaCE$ and stable mass transfer efficiency $\beta$, and the right column shows models from grid B, where we vary the supernova natal kick velocity root-mean-square $\sigmaRMS$ and the remnant mass prescription (RMP). For the RMP prescription, D, M, and R stand for the delayed, \citet{Mandel_2020}, and rapid prescription, respectively. The assumed SFRD from \citet{Madau_Fragos_2017} is scaled arbitrarily and plotted on all panels in black; note that the SFRD is, in reality, orders of magnitude greater than the merger rate but the scaling helps compare the shapes by eye. For all merger rates, we include the $1\sigma$ and $2\sigma$ confidence intervals calculated by bootstrapping the simulation results to show the sampling uncertainty. The sharp merger rate drop-off around $z \sim 9$ is due to our assumption that star formation starts at $z=10$ and mergers are delayed. 
  }
  \label{fig:all_merger_rates}
\end{figure*}

\begin{figure*}
    \centering
    \includegraphics[width=\linewidth]{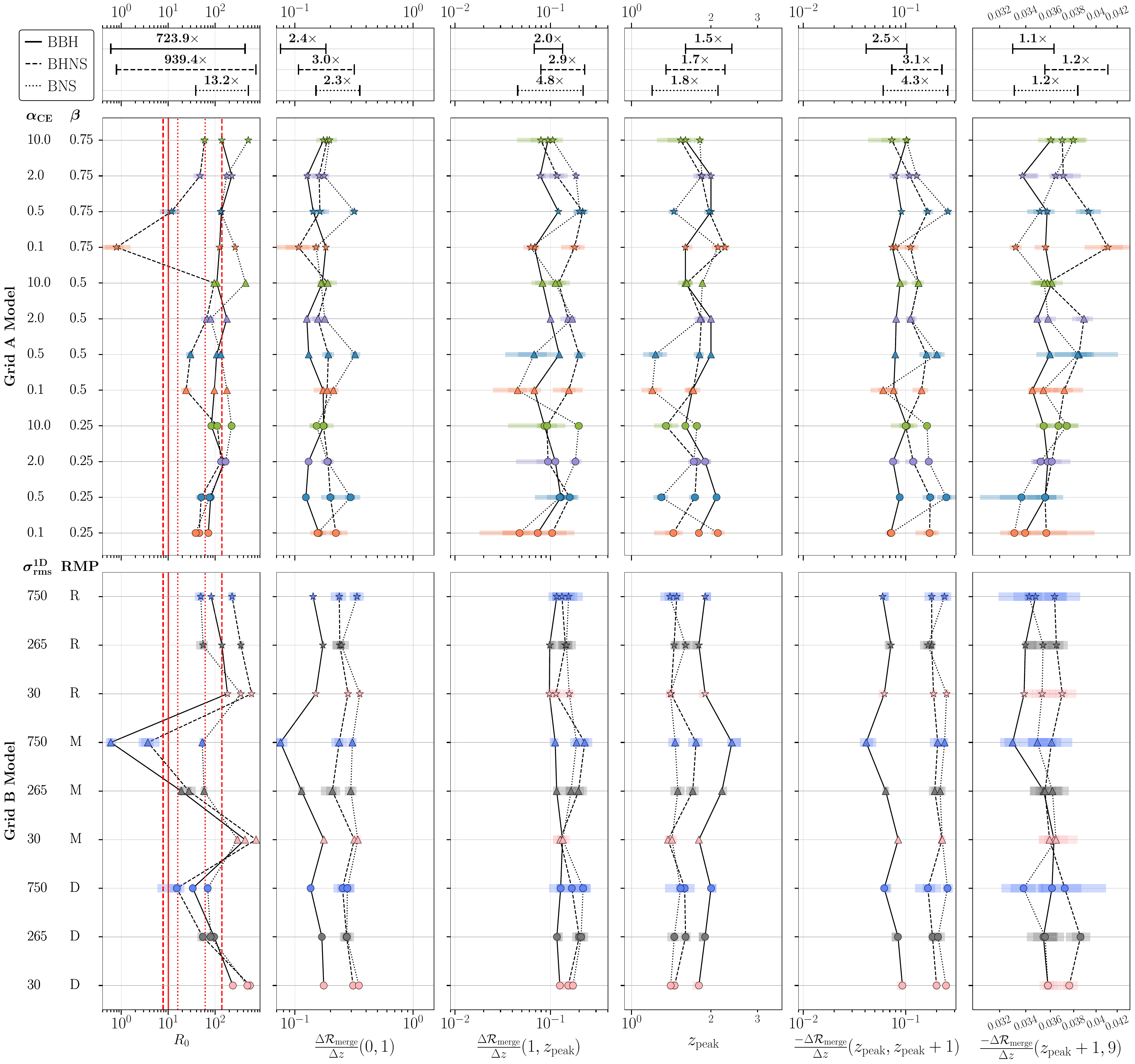}
    \caption{The intrinsic merger rate ($\mathcal{R}_0$), peak redshift ($z_{\rm{peak}}$), and merger rate differentials in the $z$ ranges $[0,1], [1,z_{\rm{peak}}], [z_{\rm{peak}}, z_{\rm{peak}}+1], [z_{\rm{peak}}+1, 9]$ for BBHs, BHNSs, BNSs.
    The top and bottom rows are for models in grid A and B, respectively, and colors and markers correspond to those in Figure~\ref{fig:all_merger_rates}.
    The red vertical lines in the first column are the inferred BBH, BHNS, and BNS local merger rates (90\% credible intervals) from \citep{Abbott:2021GWTC3}. 
    We include $1\sigma$ and $2\sigma$ confidence intervals as shaded boxes around each point, which are calculated by bootstrapping the simulation results. Along the y-axis in the left margin, RMP stands for remnant mass prescription and D, M, and R stand for the delayed, \citet{Mandel_2020}, and rapid prescriptions, respectively.}
    \label{fig:all_differentials}
\end{figure*}

We show the simulated BBH, BHNS, and BNS merger rates as a function of redshift in Figure~\ref{fig:all_merger_rates} for all binary evolution models. In
Figure~\ref{fig:all_differentials}, we show a quantitative analysis of these rates' redshift evolutions including their differential rates (defined in Equation~\ref{eq:differential}), local rates, and peak redshifts.
The differential rate is a representative of the ``normalized'' merger rate slope: values close to $1$ or $-1$ indicate steep increases or decreases whereas values close to 0 indicate flatter evolution over a given redshift interval.
We will therefore refer to these differential rates as `slopes'.
We find the following results. 

Most importantly, we observe that the BBH, BHNS, and BNS merger rates follow a remarkably similar evolution over redshift for all our models: the merger rates rise monotonically between $z=0$ until their peaks between $1.5 \lesssim z\lesssim 2.5$ and then steeply decline until $z\sim 9.5$ where they sharply drop off\footnote{The merger rate drop off around $z\sim 9.5$ in all simulations results from our assumption that star formation starts at $z=10$ combined with minimum delays of $\gtrsim 10\,{\rm{Myr}}$ between star formation and the DCO merger.}.
In Figure~\ref{fig:all_differentials} we show that the merger rates from all models peak between redshifts $1.60 < z < 2.40$ (BBH), $1.35 \lesssim z \lesssim 2.26$ (BHNS), and $1.20 \lesssim z \lesssim 2.13$ (BNS), and that their slopes typically vary with factors between $1$--$3\times$ for a given \ac{DCO} type.
The slopes that are outliers are (i) \ac{BNS} models in the range $[1, z_{\rm{peak}}]$ which vary between $0.045$ and $0.22$ (a factor $4.8\times$) and (ii) \ac{BNS} models in $[z_{\rm{peak}}, z_{\rm{peak}}+1]$, which vary between $-0.26$ and $-0.06$ (a factor $4.3\times$).
The redshift interval with the smallest variations in slope is $[z_{\rm{peak}} + 1, 9]$, for which values only span a factor of $1.1\times, 1.2\times,$ and $1.2\times$ for BBHs, BHNSs, and BNSs, respectively. 
In contrast to the similarity of the merger distributions, we find that the intrinsic merger rates can span factors of almost $1000\times$ between models, namely $724\times$, $939\times$, and $13\times$ for BBH, BHNS, and BNS, respectively\footnote{The factor $13\times$ for the BNS intrinsic merger rate variations is small and would probably also have been of order $1000\times$ if we had included more model variations (e.g., \citealt{Broekgaarden_2022, Chu:2022}). Indeed, even within our models it is clear that the magnitude of the BNS merger rate at higher redshift varies factors $\gtrsim 100\times$ between models.}.

\begin{figure*}
    \centering
    \includegraphics[width=\textwidth]{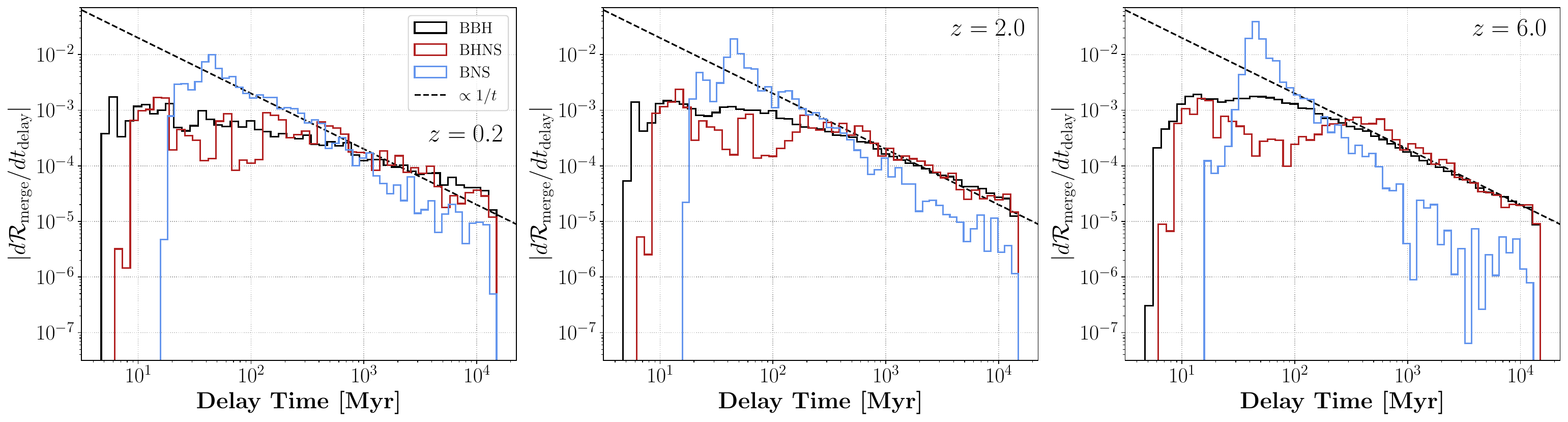}
    \caption{The delay time distribution of BBHs, BHNSs, and BNSs for the model with $\alphaCE = 2.0$ and $\beta = 0.5$. The left, middle, and right panels are the delay times for binaries formed at $z=0.2, 2.0,$ and $6.0$, respectively.}
    \label{fig:all_dco_delay_times}
\end{figure*}

The merger rate slopes are so similar between models because the two primary factors that affect the redshift distribution of binaries are relatively model-agnostic.
First, the formation efficiency as a function of metallicity follow the same general trend for a given DCO type with all of our models (see Figure~\ref{fig:all_formation_rates} for the formation rates).
Second, the delay time distribution of all models follow a $t^{-1}$--like distribution.
Combined, these two effects govern the redshift distribution of mergers because they dictate where binaries form and how long they live before merging (see \citealt{Boesky_2023_GW_paper} for more details).
In Figure~\ref{fig:all_merger_rates}, the BHNS merger rates have notably steeper slopes on average than BBHs for the range $z \gtrsim z_{\rm{peak}}$.
This is a result of BHNSs having fewer systems with short delay times ($\lesssim 1\,\rm{Gyr}$), which we show in Figure~\ref{fig:all_dco_delay_times}, leading to fewer mergers at high redshift.

%

We also notice several trends in how specific parameters impact the merger rates in Figures~\ref{fig:all_merger_rates} and \ref{fig:all_differentials}.
The dominant parameter for the redshift distribution of mergers from grid A is the common envelope efficiency $\alphaCE$, as is visible by the clustering of rates by $\alphaCE$ in Figure~\ref{fig:all_merger_rates}.
Models with $\alphaCE = 0.1, 10.0$ tend to produce fewer BBH mergers than models with $\alphaCE = 0.5, 2.0$.
Our simulations are therefore consistent earlier studies which found that $\alphaCE$ has a non-monotonic effect on binary physics \cite[e.g.][]{Broekgaarden_2022, Bavera:2022}\footnote{See \citet{Boesky_2023_GW_paper} for more details on \alphaCE.}.
Models with $\alphaCE = 0.1$ and $2.0$ produce the least BHNS mergers, whereas models with $\alphaCE = 0.5$ and $0.1$ produce the least BNSs mergers, depending on the value of $\beta$.
We also find that models with $\alphaCE = 0.1, 10.0$ tend to favor low redshift BBH mergers relative to other grid A models.

The \ac{SN} natal kick velocity dispersion $\sigmaRMS$ dominates the redshift distribution of mergers for grid B.
In Figure~\ref{fig:all_merger_rates} we find that holding the remnant mass prescription constant, the number of mergers monotonically decreases (often with more than a factor $10\times$) with increasing $\sigmaRMS$. 
This is because higher $\sigmaRMS$ values lead more binaries to disrupt during supernova.
The number of mergers span the largest range between $\sigmaRMS$ perscriptions for BBHs with the Mandel $\&$ M\"uller RMP, indicating that BBH simulations with this stochastic RMPs are particularly sensitive to $\sigmaRMS$.

\begin{figure}
    \centering
    \includegraphics[width=1\columnwidth]{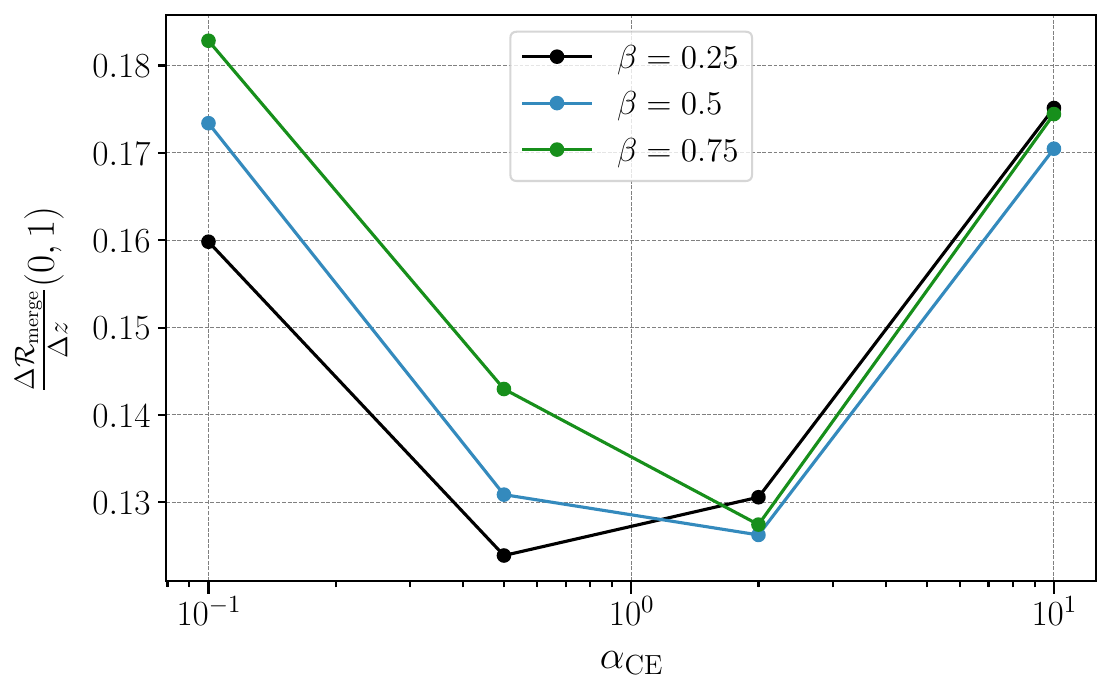}
    \caption{The differential BBH merger rate in the interval $z=0$ to $z=1$ as a function of $\alphaCE$ values for different fixed values of $\beta$ in our simualtion grid A.}
    \label{fig:diff_as_func}
\end{figure}

Trends in how parameters impact the merger rate are often consistent across values for the second grid parameter.
One example is the BBH slope as a function of $\alphaCE$ between $0 < z < 1$ with fixed $\beta$s shown in Figure~\ref{fig:diff_as_func}.
For all three $\beta$ prescriptions, the differential falls by a factor of $\sim 1/3$ from $\alphaCE = 0.1$ to $\alphaCE = 0.5, 2.0$ and then rises by a factor of $\sim 1/3$ when $\alphaCE = 10$.
In \citet{Boesky_2023_GW_paper}, we discuss how $\alphaCE$ governs delay times: large $\alphaCE$ fails to shrink orbits enough to merge in Hubble time, but small $\alphaCE$ prevents \acp{CE} from being ejected altogether, resulting in stellar mergers.
This dynamic creates a ``sweet spot'' for which the delay times of models with $\alphaCE=0.5, 2.0$ are considerably lower than those with $\alphaCE=0.1, 10.0$.
We find that models with longer delay times have larger merger rate differentials because binaries merge at lower redshift, therefore causing sharper increase from $z=0$ to $z=1$ as visible in Figure~\ref{fig:diff_as_func} \citep[cf.][]{Olejak:2022supernova}. A full overview of the parameter impacts is provided in Appendix~\ref{sec:app-parameter-impact-all}.

\section{Discussion} \label{sec:discussion}
One of our key findings is that uncertainties in massive binary stellar evolution have a relatively small impact on the merger rate slopes, but can have a significant impact on the local merger rate and overall number of mergers.
Specifically, the slopes of the merger rate are typically within a factor of $\lesssim3\times$ in a given redshift range while the intrinsic rates span factors up to $\sim 1000\times$.
These findings are in agreement with earlier work.
For example, \citet{Belczynski:2016} simulate the BBH merger rate for four different $\sigmaRMS$ models and find similar merger rate slopes between models but order of magnitude different merger rate normalizations. They also find higher $\sigmaRMS$ leading to fewer mergers, in agreement with our results in Figure~\ref{fig:all_merger_rates}.
\citet{Riley:2021} find that wind loss during the Wolf-Rayet phase does not significantly impact the BBH merger rate shape, \citet{Santoliquido_2022} find similar shapes for the merger rate as a function of redshift when varying the common-envelope efficiency, mass transfer efficiency, and natal kicks, and 
\citet{Chu:2022} find similar BNS merger rate slopes when varying the common envelope efficiency, $\sigmaRMS$, and supernova ejecta, with the exception of one model which only creates BNSs with long ($\gtrsim 1\,\rm{Gyr}$) delay times. A full review of the different merger rates predicted by different simulations is out of scope for this paper, but will be important to advance the field. 

On the other hand, our study and these works do not explore many other key uncertainties including stellar evolution tracks \citep[e.g.,][]{Agrawal:2023,Romagnolo:2023}, alternative metallicity-dependent star formation rate models \citep[e.g.,][]{Neijssel_2019, Briel:2022, Santoliquido_2022, chruslinska_2022}, and different initial stellar property distributions \citep[e.g.,][]{Klencki_2018, 2020A&A...636A..10C}.
The rapid increase in gravitational wave observations at increasing distances will improve measurements of the redshift evolution of the compact object merger rate \citep[e.g.,][]{GWTC-3_population_inference, Nitz:2023-4-OGC, Godfrey:2023, Callister:2023, Payne:2023,  Ray:2023}.
Future studies should therefore further investigate how compact object mergers are impacted by the uncertainties omitted in this study.
If the shape of the isolated binary merger rate is, however, robust across uncertainties in massive binary star evolution, it would support the potential for using the observed merger rate in tandem with simulations to constrain other uncertainties such as the star formation history and formation channel contributions.

\section{Conclusion} \label{sec:conclusion}
In this study, we presented the expected cosmological merger rates of BBHs, BHNSs, and BNSs for the isolated binary channel using population synthesis simulations generated with COMPAS. We used two two-dimensional grids of models for binary evolution to investigate the impact of stellar evolution uncertainties.
To analyze and quantify the redshift evolution of the merger rate in our simulations, we parameterized the merger rate using the rate at $z\sim 0$, the redshift of the peak merger rate ($z_{\rm peak}$), and the differentials for several redshift intervals as a proxy for the slopes (Equation~\ref{eq:differential}).
We summarize our main findings below:

\begin{enumerate}
    \item The redshift evolution of the BBH, BHNS, and BNS merger rates follow a remarkably similar shape for all our simulations (Figure~\ref{fig:all_merger_rates}): they increase from $z\sim0$ until a peak between $z=1.2$--$2.4$, and then decline until our assumed beginning of star formation at $z=10$. Although the local ($z\sim 0$) merger rate and overall normalization vary by factors up to $1000\times$ between models, the slopes (quantified  with the differential from Equation~\ref{eq:differential}) typically vary with factors of $1$--$3\times$ (Figure~\ref{fig:all_differentials}).


    \item The shape of the merger rate across redshift is correlated with specific binary evolution parameters (Figure~\ref{fig:diff_as_func}). Future observations of mergers to high redshifts can therefore help constrain models for binary evolution.

    \item We find that the common-envelope efficiency \alphaCE dominates the redshift distribution of mergers in our grid A simulations (Figure~\ref{fig:all_merger_rates}). It has a non-monotonic impact on the merger rate, which is a result of a ```sweet spot'' range in which binaries both can successfully eject their \acp{CE} and are tightened enough to merge in a Hubble time.

    \item The \ac{SN} natal kick velocity dispersion $\sigmaRMS$ typically dominates the shape of the merger rate in our grid B simulations because strong kicks disrupt the binaries leading to a drastic decrease in the efficiency of \ac{DCO} formation (Figure~\ref{fig:all_merger_rates}, Figure~\ref{fig:all_formation_rates}).

\end{enumerate}

\begin{acknowledgments}
APB acknowledges support from the Harvard PRISE and HCRP fellowships.
FSB acknowledges support for this work through the NASA FINESST scholarship 80NSSC22K1601 and from the Simons Foundation as part of the Simons Foundation Society of Fellows under award number 1141468.
\end{acknowledgments}

%

\vspace{5mm}


\software{
The simulations in this paper were performed with the COMPAS rapid binary population synthesis code version {2.31.04}, which is available for free at \url{http://github.com/TeamCOMPAS/COMPAS} \citep{COMPAS_2022}. 
The authors used {\sc{STROOPWAFEL}} from \citet{Broekgaarden_2019}, publicly available at \url{https://github.com/FloorBroekgaarden/STROOPWAFEL}.
The authors' primary programming language was \textsc{Python} from the Python Software Foundation available at \url{http://www.python.org} \citep{CS-R9526}. In addition, the following Python packages were used: \textsc{Matplotlib} \citep{2007CSE.....9...90H},  \textsc{NumPy} \citep{2020NumPy-Array}, \textsc{SciPy} \citep{2020SciPy-NMeth}, \textsc{IPython$/$Jupyter} \citep{2007CSE.....9c..21P, kluyver2016jupyter}, 
\textsc{Astropy} \citep{2018AJ....156..123A}  and   \href{https://docs.h5py.org/en/stable/}{\textsc{hdf5}} \citep{collette_python_hdf5_2014}. 
}

\bibliography{sample631}{}
\bibliographystyle{aasjournal}

\appendix

\section{Simulation Settings}

Table~\ref{tab:COMPAS_fiducial} provides a summary of our assumptions for the COMPAS population synthesis simulations.

\begin{table*}[h!]
\caption{Initial values and default settings chosen for the population synthesis simulations performed with {\sc{COMPAS}} in this study. Cyan and orange stars indicate prescriptions and assumptions that we vary in tandem (see Table~\ref{tab:COMPAS_grid}).}
\label{tab:COMPAS_fiducial}
\centering
\resizebox{\textwidth}{!}{%
\begin{tabular}{lll}
\hline  \hline
Description and name                                 														& Value/range                       & Note / setting   \\ \hline  \hline
\multicolumn{3}{c}{Initial conditions}                                                                      \\ \hline
Initial mass \monei                               															& $[5, 150]$\Msun    & \citet{Kroupa_2001} IMF  $\propto  {\monei}^{-\alpha}$  with $\alpha_{\rm{IMF}} = 2.3$ for stars above $5$\Msun	  \\
Initial mass ratio $\qi = \mtwoi / \monei $           												& $[0, 1]$                          &       We assume a flat mass ratio distribution  $p(\qi) \propto  1$ with \mtwoi $\geq 0.1\Msun$   \\
Initial semi-major axis \ai                                            											& $[0.01, 1000]$\AU & Distributed flat-in-log $p(\ai) \propto 1 / {\ai}$ \\   
Initial metallicity \Zi                                           											& $[0.0001, 0.03]$                 & Distributed flat-in-log $p(\Zi) \propto 1 / {\Zi}$        \\
Initial orbital eccentricity \ei                                 							 				& 0                                & All binaries are assumed to be circular at birth  \\
%
\hline
\multicolumn{3}{c}{Fiducial parameter settings:}                                                            \\ \hline
Stellar winds  for hydrogen rich stars                                   																&      \citet{Belczynski_2010a}    &   Based on {\citet{Vink_2000,Vink_2001}}, including  LBV wind mass loss with $f_{\rm{LBV}} = 1.5$   \\
Stellar winds for hydrogen-poor helium stars &  \citet{Belczynski_2010b} & Based on   {\citet{Hamann_1998}} and  {\citealt{Vink_deKoter_2005}}  \\

%
Max transfer stability criteria & $\zeta$-prescription & Based on \citet[][]{VignaGomez_2018} and references therein     \\ 
{\hspace{-.35cm}\Large{\textcolor{cyan}{$\star$}}}{\hspace{+.02cm}} Mass transfer accretion rate & thermal timescale & Limited by thermal timescale for stars  \citet[][]{VignaGomez_2018,Vinciguerra_2020} \\ 
 & Eddington-limited  & Accretion rate is Eddington-limit for compact objects  \\
Non-conservative mass loss & isotropic re-emission &  {\citet[][]{Massevitch_Yungelson_1975,Bhattacharya_vanDenHeuvel_1991, Soberman_1997}} \\ 
& &  {\citet{Tauris_vanDenHeuvel_2006}} \\
Case BB mass transfer stability                                														& always stable         &       Based on  \citet{Tauris_2015,Tauris_2017}, \citet{VignaGomez_2018}         \\ 
%
%
CE prescription & $\alpha-\lambda$ & Based on  \citet{Webbink_1984,deKool_1990}  \\
{\hspace{-.35cm}\Large{\textcolor{cyan}{$\star$}}}{\hspace{+.02cm}} CE efficiency $\alpha$-parameter                     												& 0.5                               &              \\
CE $\lambda$-parameter                               													& $\lambda_{\rm{Nanjing}}$                             &        Based on \citet{Xu_2010a,Xu_2010b} and  \citet{Dominik_2012}       \\
Hertzsprung gap (HG) donor in {CE}                       														& pessimistic                       &  Defined in \citet{Dominik_2012}:  HG donors don't survive a {CE}  phase        \\
%
%
{SN} natal kick magnitude \vk                          									& $[0, \infty)$\kms & Drawn from Maxwellian distribution    with standard deviation $\sigma_{\rm{rms}}^{\rm{1D}}$          \\
 {SN} natal kick polar angle $\thetak$          											& $[0, \pi]$                        & $p(\thetak) = \sin(\thetak)/2$ \\
 {SN} natal kick azimuthal angle $\phi_k$                           					  	& $[0, 2\pi]$                        & Uniform $p(\phi) = 1/ (2 \pi)$   \\
 {SN} mean anomaly of the orbit                    											&     $[0, 2\pi]$                             & Uniformly distributed  \\
{\hspace{-.35cm}\Large{\textcolor{orange}{$\star$}}}{\hspace{+.02cm}} Core-collapse  {SN} remnant mass prescription          									     &  delayed                     &  From \citet{Fryer_2012}, which  has no lower {BH} mass gap  \\%
USSN  remnant mass prescription          									     &  delayed                     &  From \citet{Fryer_2012}   \\%
ECSN  remnant mass presciption                        												&                                 $m_{\rm{f}} = 1.26\Msun$ &      Based on Equation~8 in \citet{Timmes_1996}          \\
{\hspace{-.35cm}\Large{\textcolor{orange}{$\star$}}}{\hspace{+.02cm}} Core-collapse  {SN}  velocity dispersion $\sigma_{\rm{rms}}^{\rm{1D}}$ 			& 265\kms           & 1D rms value based on              \citet{Hobbs_2005}                          \\
 USSN  and ECSN  velocity dispersion $\sigma_{\rm{rms}}^{\rm{1D}}$ 							 	& 30\kms             &            1D rms value based on e.g.,    \citet{Pfahl_2002,Podsiadlowski_2004}    \\
PISN / PPISN remnant mass prescription               											& \citet{Marchant_2019}                    &       As implemented in \citet{Stevenson_2019}      \\
Maximum NS mass                                      & $\rm{max}_{\rm{NS}} = 2.5$\Msun & Following \citet{Fryer_2012}            \\
Tides and rotation & & We do not include prescriptions for tides and/or rotation\\
\hline
\multicolumn{3}{c}{Simulation settings}                                                                     \\ \hline
Sampling method                                      & \sc{STROOPWAFEL} &                Adaptive importance sampling from  \citet{Broekgaarden_2019}.  \\
Binary fraction                                      & $f_{\rm{bin}} = 1$ &       Corrected factor to be consistent with e.g., {\citet[][]{Sana_2016}}        \\
Solar metallicity \Zsun                             & \Zsun = 0.0142 & based on {\citet{Asplund_2009}} \\
Binary population synthesis code                                      & COMPAS &       \citet{COMPAS_2022} \\
\hline \hline
\end{tabular}%
}
\end{table*}

\section{Formation rates as a function of redshift} \label{sec:formation_rates}

\begin{figure*}
  \centering
  \includegraphics[width=\linewidth]{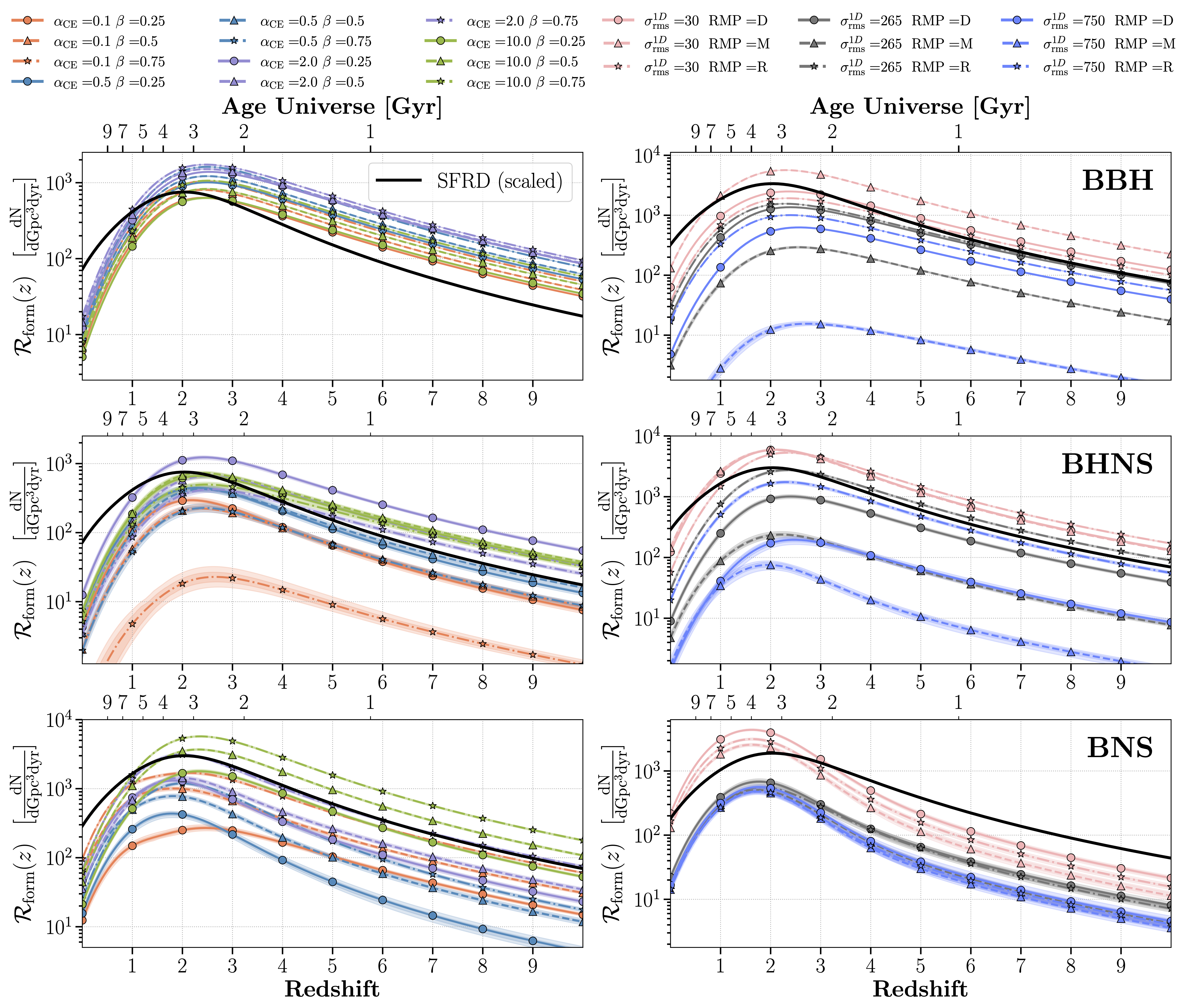}
  \caption{Same as Figure~\ref{fig:all_merger_rates}, but for the formation rates of BBH, BHNS, and BNS systems instead of merger rates. We define the formation time as directly after the second supernova when both compact objects have formed.}
  \label{fig:all_formation_rates}
\end{figure*}

Figure~\ref{fig:all_formation_rates} shows the formation rate (i.e. number of \acp{DCO} formed) as a function of redshift for all our simulations. 
It is clear that the formation rates peak at higher redshifts than the merger rates in Figure~\ref{fig:all_merger_rates} because there are non-negligible delays between binary formation and merger.
For many models the formation rate peaks at higher redshifts compared to the star formation rate peak, which is a result of boosted \ac{DCO} formation efficiency at low metallicities (see \citealt{Boesky_2023_GW_paper} for more details).

\section{Correlating Differentials With Binary Evolution Model Parameters}\label{sec:app-parameter-impact-all}

\begin{figure*}
  \centering
  \includegraphics[width=\linewidth]{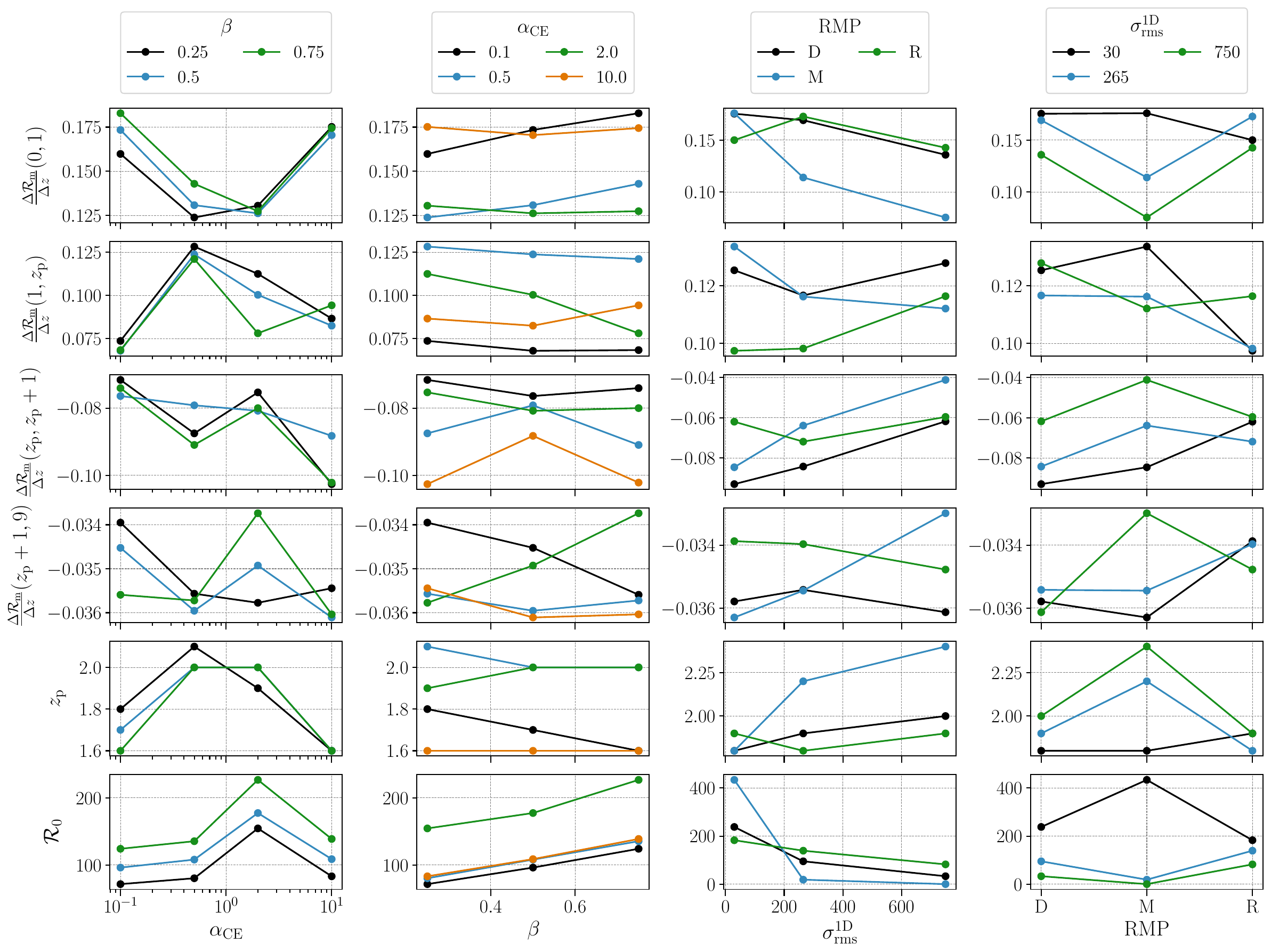}
  \caption{Differentials of the BBH merger rate in the redshift ranges $[0, 1]$, $[1, z_{\rm{peak}}]$, $[z_{\rm{peak}}, z_{\rm{peak}} + 1]$, $[z_{\rm{peak}} + 1 , 9]$ as well as $z_{\rm{peak}}$ and the local rate $\mathcal{R}_0$ plotted as a function of model parameters while keeping the second grid parameter fixed.}
  \label{fig:all_diffs_as_func_BBH}
\end{figure*}

\begin{figure*}
  \centering
  \includegraphics[width=\linewidth]{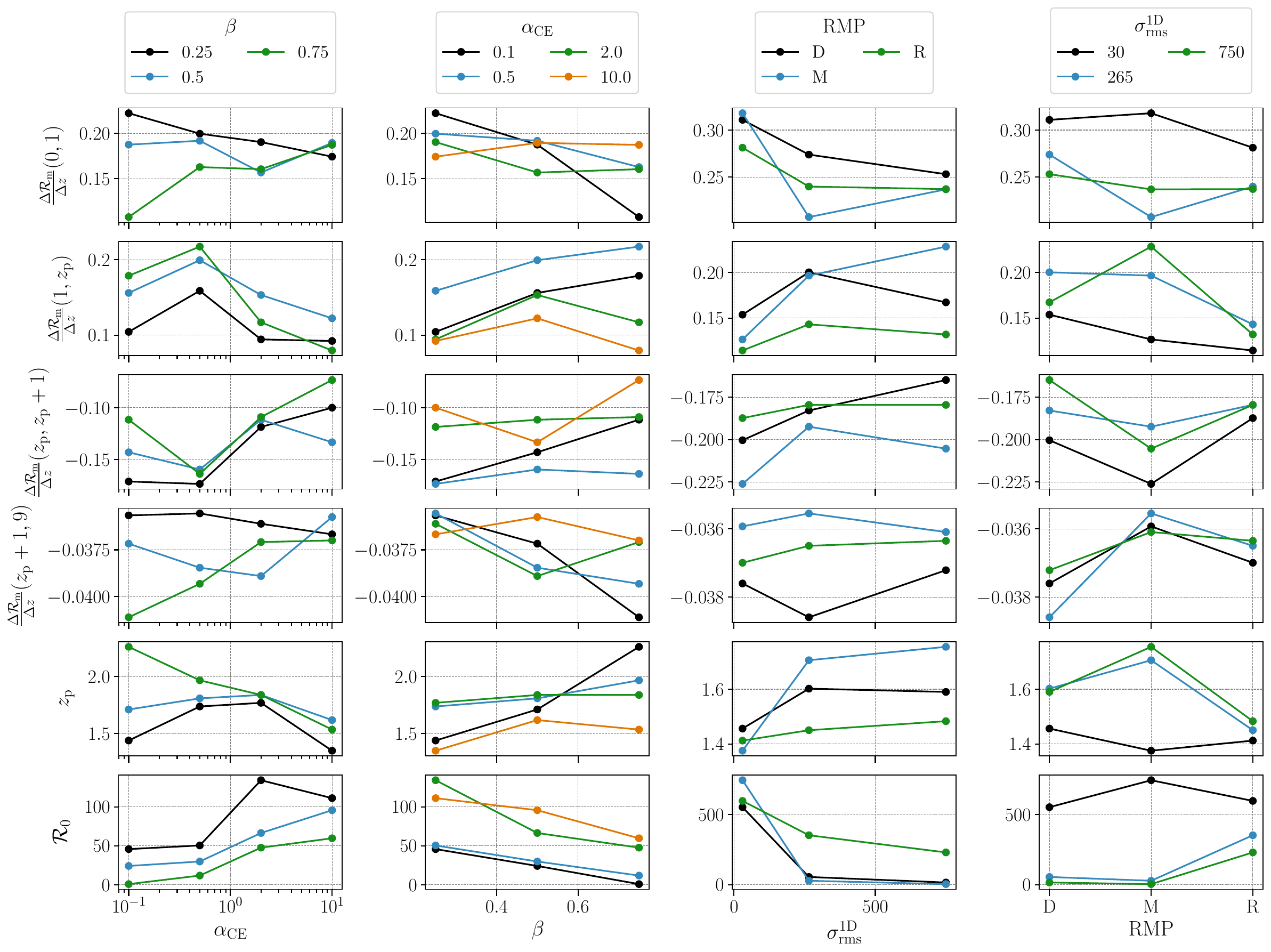}
  \caption{The same as Figure~\ref{fig:all_diffs_as_func_BBH} but for BHNSs instead of BBHs.}
  \label{fig:all_diffs_as_func_BHNS}
\end{figure*}

\begin{figure*}
  \centering
  \includegraphics[width=\linewidth]{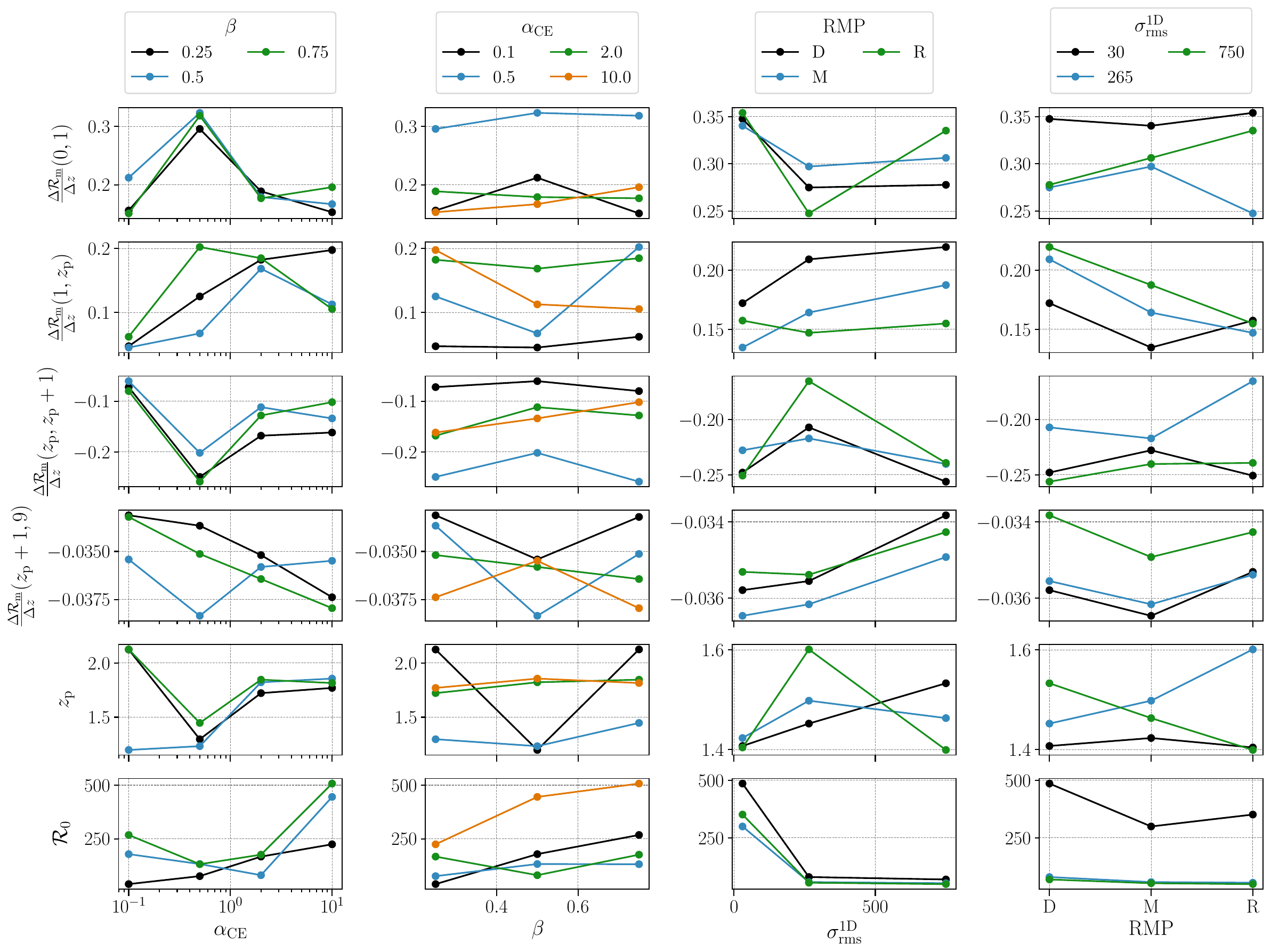}
  \caption{The same as Figure~\ref{fig:all_diffs_as_func_BBH} but for BNSs instead of BBHs.}
  \label{fig:all_diffs_as_func_BNS}
\end{figure*}

As detectors observe compact object mergers at increasing distances in the coming years, we will constrain the redshift evolution of the merger rate.
Better constraints on the redshift evolution could enable us to tune population synthesis parameters by comparing the simulated and true redshift distribution of mergers.
Parameterizing the merger rate redshift evolution with metrics such as the differential (Equation~\ref{eq:differential}) will be important for quantifying and correlating features of $\mathcal{R}_{\rm{merge}}(z)$ with model parameters.
To this end, we show the differential in several redshift ranges, $z_{\rm{peak}}$, and $\mathcal{R}_0$ as a function of model parameters for BBHs, BHNSs, and BNSs in Figure~\ref{fig:all_diffs_as_func_BBH}, Figure~\ref{fig:all_diffs_as_func_BHNS}, and Figure~\ref{fig:all_diffs_as_func_BNS}, respectively.
Besides how the BBH differentials correlate with $\alphaCE$ (described in Section~\ref{sec:results}), we leave the interpretation of trends in these figures to the reader.

\end{document}